\documentclass[aps,prb,a4paper,twocolumn]{revtex4}

\usepackage{graphicx}
\usepackage{textcomp}
\usepackage{amssymb}\usepackage{amsmath}\usepackage{amsthm}

\begin{document}
\title{Cavity QED with a Bose-Einstein condensate}
\author{Ferdinand Brennecke$^1$}
\author{Tobias Donner$^1$}
\author {Stephan Ritter$^1$}
\author{Thomas Bourdel$^2$}
\author{Michael K\"ohl$^3$}
\author{Tilman Esslinger$^1$}
\affiliation{$^1$Institute for Quantum Electronics, ETH Z{\"u}rich, 8093 Z{\"u}rich, Switzerland\\
$^2$Laboratoire Charles Fabry, Institut d'Optique, Campus Polytechnique, RD 128, F91127 Palaiseau cedex\\
$^3$Cavendish Laboratory, University of Cambridge, Cambridge CB3 0HE, United Kingdom}

\maketitle

\textbf{
Cavity quantum electrodynamics (cavity QED) describes the coherent interaction between matter and an electromagnetic field confined within a resonator structure, and is providing a useful platform for developing concepts in quantum information processing \cite{vanenk2004}. By using high-quality resonators, a strong coupling regime can be reached experimentally in which atoms coherently exchange a photon with a single light-field mode many times before dissipation sets in. This has led to fundamental studies with both microwave \cite{raimond2001,walther2002} and optical resonators \cite{kimble1998}. To meet the challenges posed by quantum state engineering \cite{mabuchi2002} and quantum information processing, recent experiments have focused on laser cooling and trapping of atoms inside an optical cavity \cite{boozer2006,nussmann2005,sauer2004}. However, the tremendous degree of control over atomic gases achieved with Bose-Einstein condensation \cite{anderson1995} has so far not been used for cavity QED. Here we achieve the strong coupling of a Bose-Einstein condensate to the quantized field of an ultrahigh-finesse optical cavity and present a measurement of its eigenenergy spectrum.
This is a conceptually new regime of cavity QED, in which all atoms occupy a single mode of a matter-wave field and couple identically to the light field, sharing a single excitation. This opens possibilities ranging from quantum communication \cite{pellizzari1995,duan2001,cirac1997} to a wealth of new phenomena that can be expected in the many-body physics of quantum gases with cavity-mediated interactions\cite{horak2000,lewenstein2006}.
}

The coherent coupling of a single two-level atom with one mode of the quantized light field leads to a splitting of the energy eigenstates of the combined system and is described by the Jaynes-Cummings model \cite{jaynes1963}. For the experimental realization the strong coupling regime has to be reached, where the maximum coupling strength $g_0$ between atom and light field is larger than both the amplitude decay rate of the excited state $\gamma$ and that of the intracavity field $\kappa$. In the case of a thermal ensemble of atoms coupled to a cavity mode, the individual, position-dependent coupling for each atom has to be taken into account.

To capture the physics of a Bose-Einstein condensate (BEC) coupled to the quantized field of a cavity, we consider $N$ atoms occupying a single wave function. Because the atoms are in the same motional quantum state, the coupling $g$ to the cavity mode is identical for all atoms. Moreover, bosonic stimulation into the macroscopically populated ground state should largely reduce the scattering of atoms into higher momentum states during the coherent evolution. This situation is therefore well described by the Tavis-Cummings model \cite{tavis1968}, where $N$ two-level atoms are assumed to identically couple to a single field mode. A single cavity photon resonantly interacting with the atoms then leads to a collective coupling of $g\sqrt{N}$.

\begin{figure}
\includegraphics[width=\columnwidth]{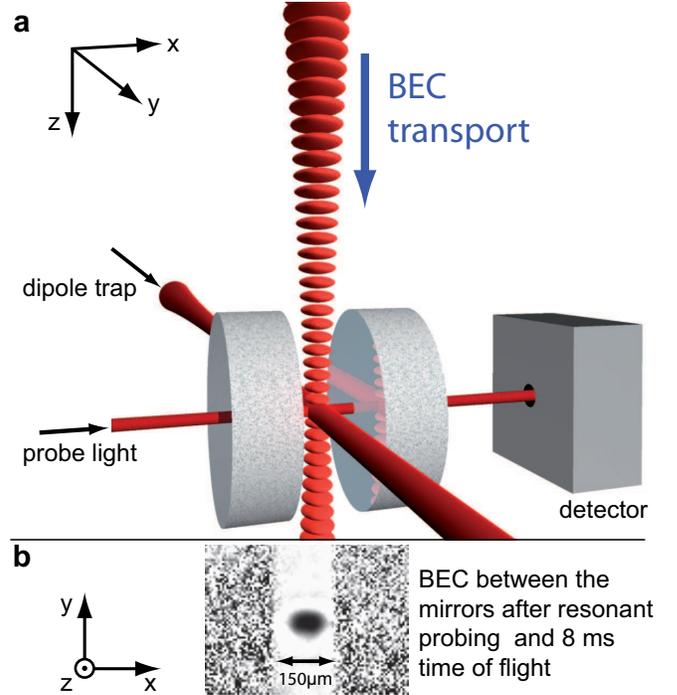}
\caption{\textbf{Experimental situation.} \textbf{a}, 36\,mm above the cavity, $3.5\times 10^6$ ultracold atoms are loaded into the dipole potential of a vertically oriented one-dimensional optical lattice. This trumpet-shaped standing wave has its waist inside the ultrahigh-finesse cavity and is composed of two counter-propagating laser beams. A translation of the lattice transports the atoms into the cavity mode. There, they are loaded into a crossed-beam dipole trap formed by a focused beam oriented along the y axis and one of the transport beams. \textbf{b}, Almost pure condensates with $2.2 \times 10^5$ atoms are obtained.}
\label{fig1}
\end{figure}

A key characteristic of the coupled BEC-cavity system is its eigenenergy spectrum, which we map out with a single excitation present.
An ensemble of thermal atoms does not fulfill the requirement of identical coupling, but it shows a similar energy spectrum, which can be modelled by the Tavis-Cummings hamiltonian with an effective collective coupling \cite{leslie2004}. In previous measurements \cite{raizen1989,tuchman2006} and also in a very recent report \cite{colombe2007}, these eigenenergies have been measured for thermal atoms coupled to a cavity. Aside from the sensitivity of the spectrum to the precise spatial distribution of the atoms, the differences between a BEC and a thermal cloud, or between a BEC and a Mott insulator, should also be accessible through the fluctuations of the coupling, that is, in the width of the resonances \cite{mekhov2007}.

First experiments bringing together BEC physics and cavities concentrated on correlation measurements using single atom counting \cite{ottl2005}, studied cavity enhanced superradiance of a BEC in a ring cavity \cite{slama2007}, observed nonlinear and heating effects for ultracold atoms in an ultrahigh-finesse cavity \cite{murch2007,gupta2007}, and achieved very high control over the condensate position within an ultrahigh-finesse cavity using atom chip technology \cite{colombe2007}.

To create a BEC inside an ultrahigh-finesse optical cavity we have modified our previous set-up \cite{ottl2006}. The experiment uses a magnetic trap 36\,mm above the cavity, where we prepare $3.5\times 10^6$ $^{87}$Rb atoms in the $|F, m_F\rangle = |1, -1\rangle$ state with a small condensate fraction present. The atoms are then loaded into the dipole potential of a vertically oriented standing wave, formed by two counter-propagating laser beams. By varying the frequency difference $\delta$ between the upwards and the downwards propagating wave, the standing-wave pattern, and
with it the confined atoms, move downwards at a velocity $v= \lambda\ \delta /2$, where $\lambda$ is the wavelength of the trapping laser \cite{kuhr2001,sauer2004}. Because of continuous evaporative cooling during the transport, the number of atoms arriving in the cavity is reduced to typically $8.4\times 10^5$ atoms with a small condensate fraction present. During the 100\,ms of transport a small magnetic field is applied to provide a quantization axis and the sample remains highly spin-polarized in the $|1,-1\rangle$ state. However, owing to off-resonant scattering in the transport beams, a small fraction of the atoms undergoes transitions into the $|F=2\rangle$ hyperfine state manifold.

At the position of the cavity mode, the atoms are loaded into a crossed-beam dipole trap formed by one of the transport beams and an additional, horizontal dipole beam with a waist radius of $w_x=w_z=27$\,\textmu m (see Fig.\,\ref{fig1}). A final stage of evaporative cooling is performed by suitably lowering the laser power to final trapping frequencies $(\omega_x, \omega_y, \omega_z) = 2 \pi \times (290, 43, 277)$\,Hz, ending up with an almost pure condensate of $2.2 \times 10^5$ atoms.

The ultrahigh-finesse cavity has a length of 176\,\textmu m and consists of symmetric mirrors with a 75\,mm radius of curvature, resulting in a mode waist radius of 25\,\textmu m. A slight birefringence splits the resonance frequency of the empty cavity for the two orthogonal, principal polarization axes by 1.7\,MHz. With the relevant cavity QED parameters $(g_0,\kappa,\gamma)=2\pi \times (10.6, 1.3, 3.0)$\,MHz, the system is in the strong coupling regime. The length of the cavity is actively stabilized using a laser at 830\,nm (ref.\,26). The intracavity intensity of the stabilization light gives rise to an additional dipole potential of $2.4\,E_\mathrm{rec}$, with the recoil energy defined as $E_\mathrm{rec}=h^2/(2m \lambda^2)$, where $m$ is the mass of the atom. The chemical potential $\mu = 1.8 E_\mathrm{rec}$ of $2.2\times 10^5$ trapped atoms is comparable to the depth of this one-dimensional lattice, so that long-range phase coherence is well established in the atomic gas. The 1/e lifetime of the atoms in the combined trap was measured to be 2.8\,s.

\begin{figure}
\includegraphics[width=\columnwidth]{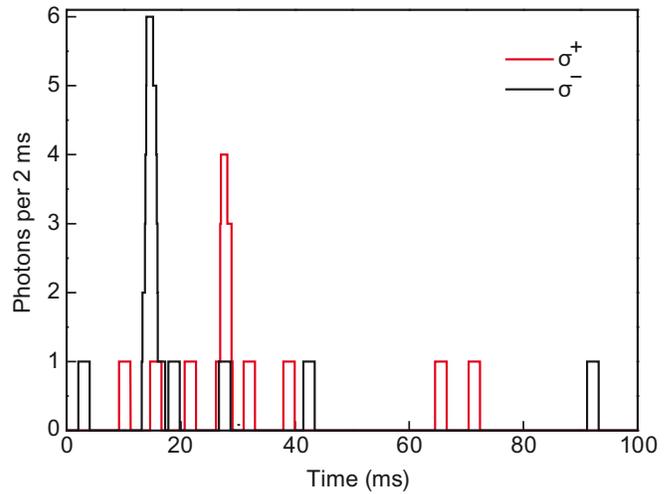}
\caption{\textbf{Cavity transmission for the $\sigma^+$ and $\sigma^-$ polarization component.} The probe laser frequency is scanned at a speed of 25\,MHz ms$^{-1}$ while the cavity detuning is fixed. The original transmission data, recorded with a resolution of 0.4\,\textmu s, is averaged over 2\,ms using a sliding average. A single peak for each polarization can clearly be distinguished from the background of about 60 dark counts per second.}
\label{fig2}
\end{figure}

To find the eigenenergies of the coupled BEC-cavity system for a single excitation, we perform transmission spectroscopy with a weak, linearly polarized probe laser of frequency $\omega_\mathrm{p}$. To this end, the resonance frequency of the empty cavity is stabilized to a frequency $\omega_\mathrm{c}$, which in general is detuned by a variable frequency $\Delta_\mathrm{c} = \omega_\mathrm{c} - \omega_\mathrm{a}$ with respect to the frequency $\omega_\mathrm{a}$ of the $| F=1 \rangle \rightarrow | F'=2 \rangle$ transition of the D$_2$ line of $^{87}$Rb. The transmission of the probe laser through the cavity is monitored as a function of its detuning $\Delta_\mathrm{p} = \omega_\mathrm{p} - \omega_\mathrm{a}$ (see Fig.\,\ref{fig2}). The two orthogonal circular polarizations of the transmitted light are separated and detected with single-photon counting modules. The overall detection efficiency for an intracavity photon is 5\,\%. To probe the system in the weak excitation limit, the probe laser intensity is adjusted such that the mean intracavity photon number is always below the critical photon number $n_0=\gamma^2/(2g_0^2)=0.04$. A magnetic field of 0.1\,G, oriented parallel (within 10\,\%) to the cavity axis provides a quantization axis.

\begin{figure}
\includegraphics[width=\columnwidth]{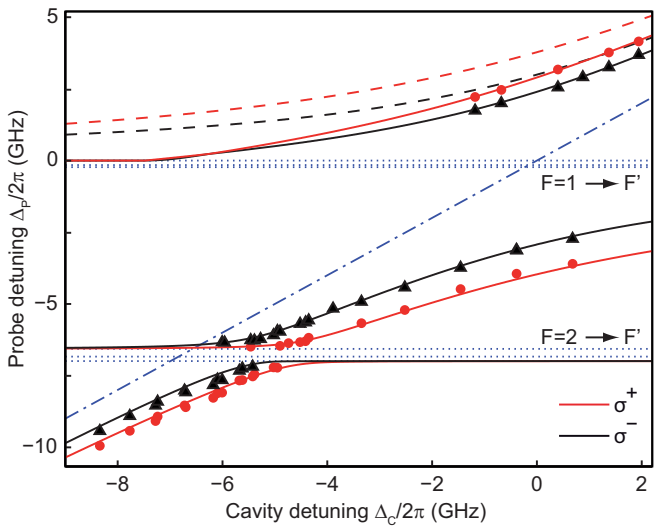}
\caption{\textbf{Energy spectrum of the coupled BEC-cavity system.} The data points are measured detunings of resonances for $\sigma^+$ (red circles) and $\sigma^-$ (black triangles) polarized light. Each data point is the average of three measurements with an uncertainty of about 25\,MHz. The solid lines are the result of a theoretical model (see Methods). Bare atomic resonances are shown as dotted lines, whereas the empty cavity resonance of the TEM$_{00}$ mode is plotted as a dashed-dotted line. Note the asymmetry in the splitting at $\Delta_\mathrm{c}=0$ caused by the influence of higher-order transverse modes. Neglecting this influence, the eigenenergies shown by the dashed lines would be expected where the free parameters were adjusted to fit the spectrum for $\Delta_\mathrm{p}<0$.}
\label{fig3}
\end{figure}

From individual recordings of the cavity transmission as shown in Fig.\,\ref{fig2} we map out the low-excitation spectrum of the coupled system as a function of $\Delta_\mathrm{c}$ (see Fig.\,\ref{fig3}). After resonant excitation we do not detect an influence on the BEC in absorption imaging for large atom numbers (see Fig.\,\ref{fig1}). For small BECs of the order of 5,000 atoms we observe a loss of 50\,\% of the atoms after resonant probing. The normal mode splitting at $\Delta_\mathrm{c}=0$ amounts to 7\,GHz for $\sigma^+$ polarization, which results in a collective cooperativity of $C = Ng^2/(2\gamma\kappa) = 1.6\times 10^6$. The splitting for the $\sigma^-$ component is smaller, because the dipole matrix elements for transitions starting in $|1,-1\rangle$ driven by this polarization are smaller than those for $\sigma^+$.

A striking feature of the energy spectrum in Fig.\,\ref{fig3} is a second avoided crossing at probe frequencies resonant with the bare atomic transitions $|F=2\rangle \rightarrow |F'=1,2,3\rangle$. It is caused by the presence of atoms in the $|F=2\rangle$ hyperfine ground state. This avoided crossing is located at a cavity detuning where the eigenenergy branch of the BEC-cavity system with no atoms in $|F=2\rangle$ would intersect the energy lines of the atomic transitions $|F=2\rangle \rightarrow |F'=1,2,3\rangle$. Accordingly, the avoided crossing is shifted by approximately $Ng^2/\Delta_\mathrm{p} = 2\pi \times 1.8$\,GHz with respect to the intersection of the empty cavity resonance with the bare atomic transition frequencies. From a theoretical analysis (see Methods), we find the size of the $|F=2\rangle$ minority component to be 1.7\,\% of the total number of atoms.

Our near-planar cavity supports higher-order transverse modes equally spaced by 18.5\,GHz, which is of the order of the collective coupling $g\sqrt{N}$ in our system. In general, the presence of one additional mode with the same coupling but detuned from the TEM$_{00}$ mode by $\Delta_\mathrm{t}$ would shift the resonance frequencies at $\Delta_\mathrm{c}$ to first order by $N g^2 / (2 \Delta_\mathrm{t})$. In our system, this results in a clearly visible change of the energy spectrum with respect to a system with a single cavity mode only (see dashed lines in Fig.\,\ref{fig3}). This can be seen as a variant of the ``superstrong coupling regime'' \cite{meiser2006}, in which the coupling between atoms and the light field is of the order of the free spectral range of the cavity.

We describe the BEC-cavity system in a fully quantized theoretical model (see Methods), which yields the eigenenergies of the coupled system. Good agreement between the measured data and the model is found (see Fig.\,\ref{fig3}) for 154,000 atoms in the $|1,-1 \rangle$ state and for 2,700 atoms distributed over the Zeeman sublevels of the $|F=2 \rangle$ state, with the majority in $|2,-1 \rangle$. The substantial influence of the higher-order transverse modes is modelled for simplicity by the coupling of the BEC to one additional effective cavity mode.

\begin{figure}[h]
\includegraphics[width=\columnwidth]{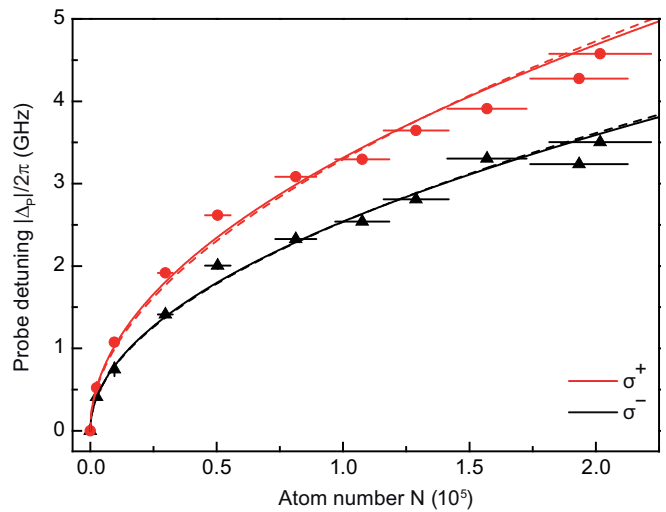}
\caption{\textbf{Shift of the lower resonance of the coupled BEC-cavity system from the bare atomic resonance.} The cavity was locked at $\Delta_\mathrm{c}=0$. $\sigma^+$ and $\sigma^-$ polarization are shown as red circles and black triangles, respectively. Each data point is the average of three measurements. The atom number was determined separately from absorption images with an assumed error of $\pm$10\%; the vertical error bars are too small to be resolved. The dashed lines are fits of the square root dependence on the atom number, as predicted by the Tavis-Cummings model. The solid lines are fits of a more detailed theoretical model (see Methods) resulting in maximum coupling rates
$g_{\sigma^+}=2\pi \times (14.4\pm0.3)$\,MHz and $g_{\sigma^-}=2\pi \times (11.3\pm0.2)$\,MHz. The ratio of the two coupling rates of $1.27\pm 0.03$ agrees with the expected ratio of $1.29$ of the corresponding Clebsch-Gordan coefficients.}
\label{fig4}
\end{figure}

To test the square-root dependence of the normal mode splitting on the number of atoms in the BEC, a second measurement was conducted. We set the cavity frequency to $\Delta_\mathrm{c}=0$ and record the detuning of the lower coupled state from the bare atomic resonance $| F=1 \rangle \rightarrow | F'=2 \rangle$ as a function of the number of atoms as displayed in Fig.\,\ref{fig4}. The atom number was varied between 2,500 and 200,000, determined from separately taken absorption images with an estimated statistical error of $\pm 10$\,\%; possible systematic shifts are estimated to be within $\pm 7$\,\%. The dependence of $|\Delta_\mathrm{p}|$ on the number of atoms is well described by a square root, as expected from the Tavis-Cummings model (dashed lines). However, for a weakly interacting BEC the size of the atomic cloud---and thus the spatial overlap with the cavity mode---depends on the atom number. Our more detailed model, which includes this effect, as well as the influence of higher-order cavity modes, yields maximum single-atom couplings of $g_{\sigma^+}=2\pi \times (14.4\pm0.3)$\,MHz and $g_{\sigma^-}=2\pi \times (11.3\pm0.2)$\,MHz for the two polarization components (solid lines). The ratio of these two couplings is $1.27\pm 0.03$ and agrees with the ratio of $1.29$ that is obtained from the effective Clebsch-Gordan coefficients for the $\sigma^+$ and $\sigma^-$ transitions starting in state $|1,-1 \rangle$.

The coupling of a single mode of a matter-wave field to a single cavity mode opens a route to new experiments. It facilitates the manipulation and study of statistical properties of quantum-degenerate matter-wave fields by a quantized optical field, or even the generation of entanglement between these two fields \cite{mekhov2007,moore1999a}. The detection of single atoms falling through the cavity has already been demonstrated with this set-up \cite{ottl2005}. In principle, the detection of small impurity components embedded in a large BEC presented here can also be extended to single atoms. This is an important step towards the realization of schemes aiming at the cooling of qubits immersed in a large BEC \cite{daley2004}.\\

\noindent\textbf{METHODS SUMMARY}

\noindent\textbf{Optical transport.} The transport of the atoms into the cavity is accomplished in $T=100$\,ms with a maximum acceleration of $a=22.4$\,m\,s$^{-2}$. The standing wave used to transport the atoms has its waist $(w_x, w_y)=(25, 50)$\,\textmu m centred inside the cavity and is locked to a $^{133}$Cs resonance at a wavelength of 852\,nm. The intensity and frequency of each beam are precisely controlled by acousto-optical modulators, which are driven by two phase-locked, homebuilt direct digital synthesis generators. The frequency difference $\delta$ between the two counterpropagating waves follows $\delta(t) = [1-\cos{(2 \pi t/T)}] \delta_\mathrm{max}/2$, with a maximum detuning of $\delta_\mathrm{max} = 1,670$\,kHz. With the maximally available power of 76\,mW per beam the trap depth at the position of the magnetic trap is 1.1\,\textmu K. During transport, the power in the laser beams is kept constant until the intensity at the position of the atoms has increased by a factor of ten. Subsequently, this intensity is kept constant.

\noindent\textbf{Theoretical model.} To gain understanding of the presented measurements we have developed a fully quantized theoretical description of the coupled BEC-cavity system. Our model includes all Zeeman sublevels in the $5 ^2\mathrm{S}_{1/2}$ and $5 ^2\mathrm{P}_{3/2}$ state manifolds of $^{87}\mathrm{Rb}$ and both orthogonal polarizations of the $\mathrm{TEM}_{00}$ cavity mode. For simplicity, the effect of higher-order transverse cavity modes is modelled by the effective coupling to one additional mode which is detuned from the $\mathrm{TEM}_{00}$ mode by $\Delta_\mathrm{t} = 2\pi \times 18.5\,\mathrm{GHz}$. The free parameters of the model are the coupling strength between this effective mode and the BEC and the population of the several ground states in the condensate. Good agreement with the measured energy spectrum is found for the ground-state population given in the main text and a coupling between BEC and effective cavity mode that is $r=1.2$ times the coupling to the $\mathrm{TEM}_{00}$ mode.\\

\noindent\textbf{METHODS}\\
To characterize the coupled BEC-cavity system theoretically, we start with the second-quantized hamiltonian describing the matter-light interaction in the electric-dipole and rotating-wave approximation \cite{moore1999a}. We take all hyperfine states including their Zeeman sublevels in the $5 ^2\mathrm{S}_{1/2}$ and $5 ^2\mathrm{P}_{3/2}$ state manifolds of $^{87}\mathrm{Rb}$ into account and describe the cavity degrees of freedom by two orthogonal linear polarizations of the $\mathrm{TEM}_{00}$ mode. We choose the quantization axis, experimentally provided by a small magnetic field, to be oriented parallel to the cavity axis. The near-planar cavity supports higher-order transverse modes equally spaced by $\Delta_\mathrm{t}=2\pi \times 18.5\,\mathrm{GHz}$. To incorporate the coupling to all higher-order transverse modes we include one additional effective cavity mode in our model with its resonance frequency shifted by $\Delta_\mathrm{t}$ with respect to that of the $\mathrm{TEM}_{00}$ mode.

Considering only one spatial atomic mode for the ground state manifold and another one for the excited-state manifold, the hamiltonian of the uncoupled system reads
\begin{equation*}
\hat{H}_0=\sum_{i}\hbar \omega_{g_{i}} \,\hat{g}_{i}^\dagger \hat{g}_{i}+\sum_{j}\hbar \omega_{e_{j}} \,\hat{e}_{j}^\dagger \hat{e}_{j}+\sum_{k=0}^1\sum_{p=\rightarrow,\uparrow}\hbar \omega_k \,\hat{a}_{k,p}^\dagger \hat{a}_{k,p}
\end{equation*}
where the indices $i$ and $j$ label the states $5 ^2\mathrm{S}_{1/2}|F, m_F\rangle$ and $5 ^2\mathrm{P}_{3/2}|F', m_{F'}\rangle$, respectively. The operators $\hat{g}^\dagger_i$ (or $\hat{g}_i$) and $\hat{e}^\dagger_j$ (or $\hat{e}_j$) create (or annihilate) an atom in the mode of the corresponding ground and excited states with frequencies $\omega_{g_{i}}$ and $\omega_{e_{j}}$. The operators $\hat{a}_{k,p}^\dagger$ (or $\hat{a}_{k,p}$) create (or annihilate) a photon with energy $\hbar \omega_k$ and linear polarization $p$ in the cavity mode $k$, where $k=0,1$ labels the $\mathrm{TEM}_{00}$ mode and the additional effective cavity mode, respectively.

The coupling between the BEC and the cavity is described by the interaction hamiltonian
\begin{equation*}
\hat{H}_\mathrm{int}=-i \hbar \sum_{k=0}^1\sum_{p=\rightarrow,\uparrow}\sum_{i,j}g_{ij}^{k,p}\, \hat{e}_{j}^\dagger \,\hat{a}_{k,p}\,\hat{g}_i+\mathrm{h.c.}
\end{equation*}
where $g_{ij}^{k,p}$ denotes the coupling strength for the transition $i\rightarrow j$ driven by the cavity mode $k$ with polarization $p$, and h.c. is the hermitian conjugate. For the $\mathrm{TEM}_{00}$ mode the coupling strength $g_{ij}^{0,p}$ depends on the dipole matrix element $\mathcal{D}_{ij}^p$ for the transition $i\rightarrow j$ driven by the polarization $p$, the mode volume $V_0$, and the overlap $\mathcal{U}_0$ between the two spatial atomic modes and the $\mathrm{TEM}_{00}$ mode:
\begin{equation*}
g_{ij}^{0,p}=\mathcal{D}_{ij}^p\sqrt{\frac{\hbar \omega_0}{2 \epsilon_0 V_0}} \,\mathcal{U}_0.
\end{equation*}
We numerically calculated the ground state of the Gross-Pitaevskii equation in the potential formed by the dipole trap and the cavity stabilization light for $N = 2 \times 10^5$ atoms, and found the spatial overlap between BEC and $\mathrm{TEM}_{00}$ mode to be $\mathcal{U}_0=0.63$. Because of the position uncertainty of the BEC relative to the cavity mode, the overlap might deviate by up to 20\,\% from this value. The repulsive interaction between the atoms leads to a slight decrease in $\mathcal{U}_0$ with increasing atom number, which was found numerically to follow $\mathcal{U}_0(N) = \sqrt{0.5}(1-0.0017 N^{0.34})$. For the coupling strength to the additional effective cavity mode we assume $g_{ij}^{1,p} = r g_{ij}^{0,p}$, with $r$ being a free parameter.

The initial atomic population of the several ground states $i$ given by the free parameters $N_i$ is relevant for the form of the energy spectrum of the coupled BEC-cavity system. To find the energy spectrum in the weak excitation limit we can restrict the analysis to states containing a single excitation. These are the states $|1_{k,p}; N_1, \ldots, N_8;0\rangle$, where one photon with polarization $p$ is present in the mode $k$ and all atoms are in the ground-state manifold, and also all possible states $|0_{k,p}; N_1, \ldots, N_i-1,\ldots, N_8;1_j\rangle$, where no photon is present and one atom was transferred from the ground state $i$ to the excited state $j$. Diagonalization of the hamiltonian $\hat{H}=\hat{H}_0+\hat{H}_{\mathrm{int}}$ in the truncated Hilbert space spanned by these states yields the eigenspectrum of the coupled system as a function of the cavity detuning $\Delta_\mathrm{c}$. The relevant transition energies with respect to the energy of the initial ground state are plotted in Fig.\,\ref{fig3}.

The detuning of the lower resonance branch at $\Delta_\mathrm{c}=0$ is a function of the atom number and given by
\begin{equation*}
|\Delta_\mathrm{p}| = \mathcal{U}_0(N) g_{\sigma^\pm} \sqrt{N} + \frac{(\mathcal{U}_0(N) r g_{\sigma^\pm})^2N}{2 \Delta_\mathrm{t}}+ \mathcal{O}(1/\Delta_\mathrm{t}^2)
\end{equation*}
Fitting this dependence to the data in Fig.\,\ref{fig4} yields the maximum coupling strengths $g_{\sigma^\pm}$ for the two polarization components $\sigma^\pm$.

\setlength{\parindent}{0pt}
\textbf{Acknowledgements} We thank A. \"Ottl for early contributions to the experiment, R. J\"ordens and A. Frank for developing the direct digital synthesis generators used for the optical transport, H. Ritsch and A. Imamoglu for discussions and OLAQUI and QSIT for funding. T.\,B. acknowledges funding by an EU Marie Curie fellowship.

\textbf{Author Information} Correspondence should be addressed to T.\,E. (esslinger@phys.ethz.ch).

\end{document}